\documentclass[preprint,prl,superscriptaddress,groupedaddress]{revtex4}

\usepackage{verbatim}
\usepackage{textcomp}
\usepackage[official]{eurosym}
\usepackage{amsmath}
\usepackage{amssymb}
\usepackage{bm}

\usepackage{tikz}
\usetikzlibrary{arrows}
\usepackage{xcolor}
\usepackage{lipsum}

\newcommand{\xv}{\bm{{\rm x}}}
\newcommand{\Rv}{\bm{{\rm R}}}
\newcommand{\Av}{\bm{{\rm A}}}

\newcommand{\uv}{\bm{{\rm u}}}

\newcommand{\Wv}{\bm{{\rm W}}}

\newcommand{\ev}{\bm{{\rm e}}}

\newcommand{\kv}{\bm{{\rm k}}}
\newcommand{\Fv}{\bm{{\rm F}}}
\newcommand{\av}{\bm{{\rm a}}}
\newcommand{\delv}{\bm{{\rm \delta}}}

\newcommand{\rv}{\bm{{\rm r}}}

\newcommand{\qv}{{\bm{{\rm q}}}}

\newcommand{\Uv}{\bm {{\rm U}}}

\begin{document}

\widetext

\title{Sonic Landau-level lasing and synthetic gauge fields in mechanical metamaterials}
\author{Hamed Abbaszadeh}
\affiliation{Instituut-Lorentz, Universiteit Leiden, Leiden 2300 RA, The Netherlands}
\author{Anton Souslov}
\affiliation{Instituut-Lorentz, Universiteit Leiden, Leiden 2300 RA, The Netherlands}
\author{Jayson Paulose}
\affiliation{Instituut-Lorentz, Universiteit Leiden, Leiden 2300 RA, The Netherlands}
\author{Henning Schomerus}
\affiliation{Department of Physics, Lancaster University, Lancaster LA1 4YB, United Kingdom}
\author{Vincenzo Vitelli}
\affiliation{Instituut-Lorentz, Universiteit Leiden, Leiden 2300 RA, The Netherlands}

\begin{abstract}
\noindent Mechanical strain can lead to a synthetic gauge field that controls the dynamics of electrons in graphene sheets as well as light in photonic crystals. Here, we show how to engineer an analogous synthetic gauge field for lattice vibrations. Our approach relies on one of two strategies: shearing a honeycomb lattice of masses and springs or patterning its local material stiffness. As a result, vibrational spectra with discrete Landau levels are generated. Upon tuning the strength of the gauge field, we can control the density of states and transverse spatial confinement of sound in the metamaterial. We also use the gauge field to design waveguides in which sound propagates robustly, as a consequence of the change in topological polarization that occurs along a domain wall in the bulk of the metamaterial. By introducing dissipation, we can selectively enhance the domain-wall-bound topological sound mode, a feature that may be exploited for the design of sound amplification by stimulated emission of radiation -- SASERs, the mechanical analogs of lasers.
\end{abstract}

\date{\today}
\maketitle

Electronic systems subject to a uniform magnetic field experience a wealth of fascinating phenomena such as topological states~\cite{Kane2005} in the integer quantum Hall effect~\cite{Hasan2010} and anyons associated with the fractional quantum Hall effect~\cite{Laughlin1983}.
Recently, it has been shown that in a strained graphene sheet, electrons experience external potentials that can mimic the effects of a magnetic field, which results in the formation of Landau levels and edge states~\cite{Guinea2010, Levy2010}. 
Working in direct analogy with this electronic setting, pseudo-magnetic fields 
have been engineered by arranging CO molecules on a
gold surface~\cite{Gomes2012} and in photonic honeycomb-lattice metamaterials~\cite{Rechtsman2013, Schomerus2013}.

In this article, we apply insights about wave propagation in the presence of a gauge field to acoustic phenomena in a nonuniform phononic crystal,
using the appropriate mechanisms of strain-phonon coupling and frictional dissipation, in contrast to those present in electronic and photonic cases. 
We develop two strategies for realizing a uniform pseudo-magnetic field in a metamaterial based on the honeycomb lattice, i.e., ``mechanical graphene''~\cite{Kariyado2015}.
In the first strategy, we apply stress at the boundary to obtain nonuniform strain in the bulk, which leads to a Landau-level spectrum, whereas in 
the second strategy, we exploit built-in, nonuniform patterning of the local metamaterial stiffness.

We explore acoustic phenomena associated with the Landau-level spectrum. 
For example, the acoustic analog of Shubnikov-de Haas oscillations~\cite{Shubnikov1930} corresponds to a sharp peak in the phonon density of states at the Landau-level frequency. 
In addition, sound modes are confined within a length scale set by the analog of the magnetic length.
Even stronger confinement of sound modes can be engineered at a domain wall associated with a change in the effective mass of the phononic excitations, which localizes phonon modes that are analogous to the topological domain-wall states in the Su-Schrieffer-Heeger model of polyacetylene~\cite{Su1979}. 
Like other realizations of topological states~\cite{Prodan2009,Berg2011} in mechanical~\cite{Kane2014,Po2014,Chen2014,Nash2015,Paulose2015,Paulose2015b,Rocklin2015,Rocklin2015a,Susstrunk2015, Chen2016,Meeussen2016}, acoustic~\cite{Khanikaev2015,Yang2015,Deymier2015, Mousavi2015,Wang2015a,Wang2015b,Xiao2015,Susstrunk2016}, and photonic~\cite{Lu2014} metamaterials, this characterization may help with the design of robust devices.
Introducing dissipation on just one of the two sublattices enhances the domain-wall-bound sound mode. This feature may be exploited for the design of acoustic couplers, rectifiers, and sound amplification by stimulated emission of radiation (SASERs).

\begin{figure}
	\includegraphics[angle=0]{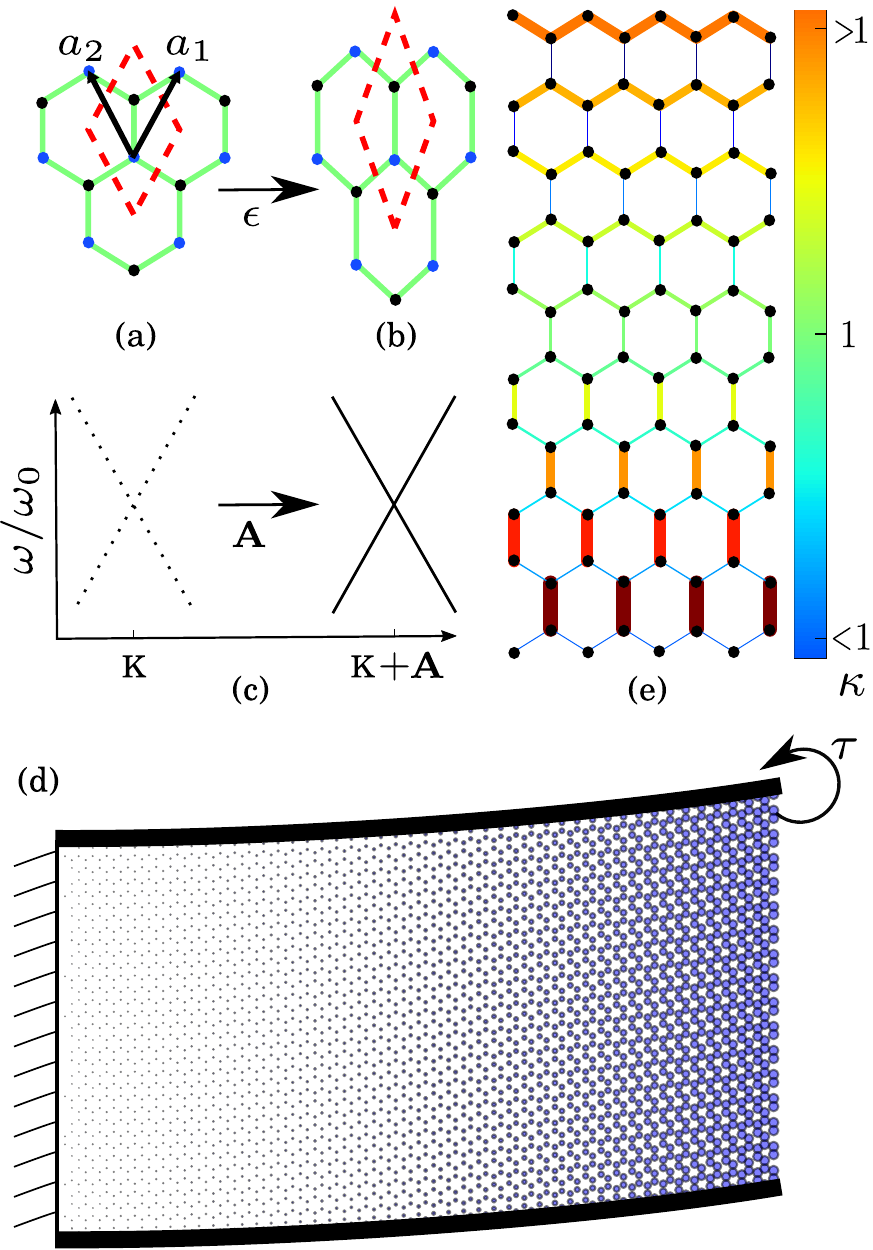}
	\caption{(a) Mechanical graphene -- a set of rods and nodes based on the honeycomb structure. The dashed line indicates the shape of a unit cell. (b) The lattice with a pure shear strain. (c) The shift of a Dirac point within the phonon spectrum of mechanical graphene due to the applied strain can be used to define an effective vector potential. (d) An externally applied nonuniform pure shear deformation that corresponds to a constant magnetic field. The external stress is applied by a torque $\tau$ on the boundary rods. (e) A non-uniform patterning of the local material stiffness that leads to a constant magnetic field. We consider periodic boundary conditions along $ x $ and free boundary conditions along $ y $.} 
	\label{Fig1}
\end{figure}

{\it Mechanical graphene.}--- We begin with a minimal, microscopic model of an acoustic metamaterial -- a set of nodes positioned at the vertices of a honeycomb lattice and connected by rods to their nearest neighbors (see Fig.~\ref{Fig1}a)~\cite{Kariyado2015}. The compressional stiffness of the rods is determined by their fixed Young's modulus and variable cross-section. We assume the rods to be so slender that
their bending stiffness is significantly lower than their compressional stiffness. 
We model the rods as central-force harmonic springs of rest length $a$, whose elastic energy $U$ is given in terms of the strain $\delta r/a$ by 
$U(\delta r) = \frac{1}{2} \kappa \left( |\rv+ \delta \rv| - a \right)^2 $. For small strains, this energy can be linearized in terms of node displacements 
$\uv_1$ and $\uv_2$ as $U(\uv_1,\uv_2) = \frac{1}{2} \kappa \left( \ev \cdot [\uv_1 - \uv_2] \right)^2 $, where $\ev \equiv \rv / |\rv|$ is the unit vector along the spring.
(In Fig.~\ref{Fig1}a, we define the initial configuration for the node positions and stiffnesses.)
Given this potential, we write down and solve the linear equation of motion for acoustic vibrations of the lattice: 
\begin{equation}
\label{eq:motion}
-m \ddot{u}^{\alpha}_i = \frac{\partial U}{ \partial u^\alpha_i} = \sum_{j,\beta} D^{\alpha\beta}_{ij} u^\beta_j,
\end{equation}
where $u^\alpha_i$ are the $\alpha=x,y$ components of displacement of the $i$th site and $D^{\alpha \beta}_{ij}$ are components of the dynamical matrix.
In a periodic lattice, the solutions to this equation of motion are plane waves $\uv_\qv e^{i (\omega(\qv) t - \qv\cdot \xv)}$,
where both the dispersion relation $\omega(\qv)$ and the normal modes $\uv_\qv$ are found from the corresponding eigenvalue problem for each wavevector $\qv$.

To lowest order in perturbation theory around point K [defined by $\qv_K \equiv (0,4 \pi / 3 \sqrt{3} a)$], the dynamical matrix for the two bands near the frequency $\omega_0 \equiv \sqrt{3 k/2 m}$ reduces to~\cite{F1}
\begin{equation}
\label{eq:H}
D = \frac{3\kappa}{4m}(a\delta\qv+\Av)\cdot {\bm{{\rm \sigma}}} + (1+V) \omega^2_0 \mathbb{I},
\end{equation}
where $\mathbb I$ is the $2\times2$ identity matrix, $\delta \qv \equiv \qv - \qv_K$, and $\bm{{\rm \sigma}} \equiv (\sigma_x, \sigma_y)$, where
\begin{equation}
\sigma_x =\left(\begin{array}{cc}
0&1 \\
1&0 \end{array}\right); \,\,\,\,
\sigma_y =\left(\begin{array}{cc}
0&-i \\
i&0 \end{array}\right)
\end{equation}
are the Pauli spin matrices. The gauge field $\Av$ and the potential $V$ are both zero for the homogeneous honeycomb lattice. 
From the structure of Eq.~(\ref{eq:H}), we note that the dispersion around $\qv_K$ has the form of a Dirac cone, i.e., the two bands touch at the Dirac point~\cite{SI}. 

{\it Synthetic gauge field.}--- We now proceed to show that unlike uniform lattice deformations that merely shift this Dirac cone in wavevector space, nonuniform deformations can lead to an effective synthetic gauge field for sound.
For uniform strain (Fig.~\ref{Fig1}b), $\Av$ and $V$ are both constant throughout the lattice.  On the other hand, for a nonuniform but slowly varying strain, the position of the local Dirac point varies from one region to another (Fig.~\ref{Fig1}c), which corresponds to fields $\Av$ and $V$ that depend on spatial coordinates. In terms of the affine component $\Uv$ and nonaffine component $\Wv$ of the displacement denoting, respectively, the common and relative displacements of the two sublattices,
\begin{align}
\label{eq:eps1}
\Av(x,y; \epsilon, W)=&  a(\qv_K\cdot\nabla)\Uv+\left[ \frac{3}{2}(\epsilon_{xx}-\epsilon_{yy}),-3\epsilon_{xy})\right] \nonumber \\
& + ( W_y, - W_x)/a,
\end{align}
and $V = \frac{1}{2} \, \mathrm{Tr} \, \epsilon$, where $\epsilon_{ij} \equiv (\partial_i U_j+\partial_j U_i)/2$ is the linear affine strain.

To simplify the design of an acoustic device based on this strained lattice, we now consider those lattice strains that can be obtained by applying forces only on the boundary. Such a configuration requires that the forces in the bulk of the material balance each other. In the material we consider, this force-balance condition is satisfied provided that the nonaffine displacements depend on the affine strain via $ W_x=\epsilon_{xy}a$ and $ W_y=\frac{1}{2}(\epsilon_{xx}-\epsilon_{yy})a$. Thus, we obtain the following expression for the gauge field in a boundary-strained material by substituting these nonaffine displacements into Eq.~(\ref{eq:eps1}):
\begin{equation}
\label{eq:eps2}
\Av(x,y; \epsilon) = a(\qv_K\cdot\nabla)\Uv+\left[ 2(\epsilon_{xx}-\epsilon_{yy}),-4\epsilon_{xy})\right].
\end{equation}

For acoustic systems, we can also follow a second strategy: patterning the local material stiffness to achieve a spatially dependent gauge field $\Av$.
For example, we can smoothly vary the composition or thickness of the rods to change their effective spring constants to $\kappa_i = \kappa +\delta \kappa_i$, where $i=1\ldots3$ labels springs in the lattice unit cell.
In this case, we find that the gauge field and potential are given by
\begin{align}
\label{eq:kappa}
\Av(x,y; \delta \kappa) &= \left (-\frac{1}{3} \frac{2 \delta \kappa _1 + \delta \kappa_2+ \delta \kappa_3}{\kappa}
, \frac{\delta \kappa_2- \delta \kappa_3}{\sqrt{3} \kappa} \right), \nonumber \\
V &= \frac{\delta \kappa_1 + \delta \kappa_2 + \delta \kappa_3}{3\kappa}.
\end{align}

\begin{figure*}
	\includegraphics[angle=0]{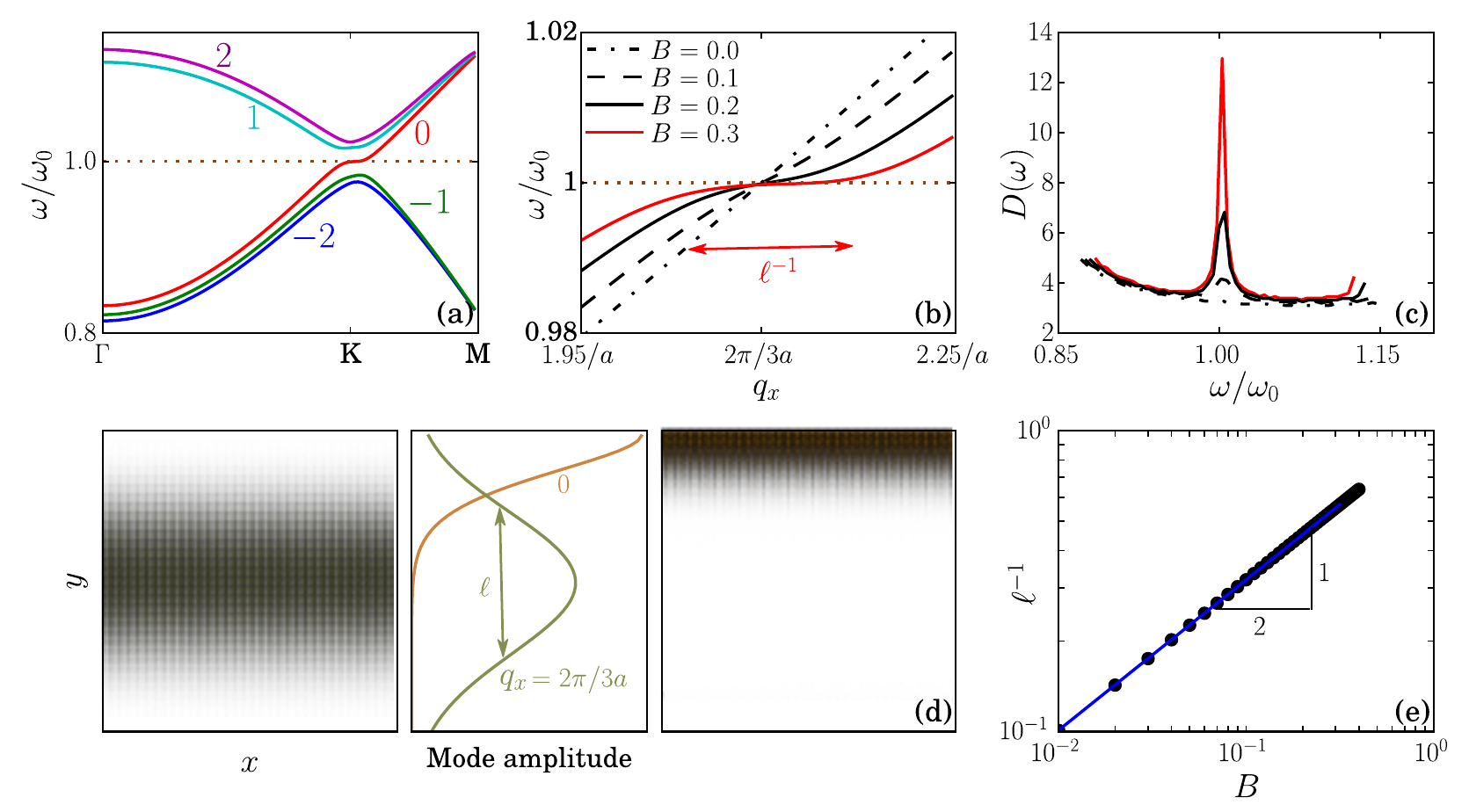}
	\caption{ Mechanical Landau levels: (a) A pseudo-magnetic field leads to Landau levels around the Dirac point. (b) As the magnetic field increases, the zeroth-Landau-level band flattens. Band flatness can be characterized by the inverse magnetic length $\ell^{-1}$. (c) Density of states for the zeroth Landau level, for the same values of $B$ as in (b). The peak at the Dirac frequency rises as the bands flatten. (d) Visualizations of the zeroth Landau level at two different wavevectors. For  $ \qv = \qv_{K}$, this mode has a Gaussian profile around the waveguide center, whereas far from this point, at $ \qv = 0$, the mode decays exponentially away from the edge. (e) The inverse magnetic length scales as the square root of the magnetic field.}
	\label{Fig2}
\end{figure*}

To obtain a Landau-level spectrum, we select $\Av$ and $V$ such that (for units in which a = 1)
\begin{equation}
\label{eq:av}
\nabla \times \Av = B \hat{z} = \textrm{const}; \,\, V = 0.
\end{equation}
For any selection satisfying the conditions of Eqs.~(\ref{eq:av}), the dynamical matrix in Eq.~(\ref{eq:H}) has the form of the Hamiltonian for a Dirac electron in a plane with a constant magnetic field $B$ applied perpendicular to that plane~\cite{Jackiw1984, Semenoff1984}. 
Let us now consider two practical solutions to Eqs.~(\ref{eq:av}):  \emph{(i)} an externally applied nonuniform pure shear deformation, and  \emph{(ii)} nonuniform patterning of the spring constants along the $y$-direction.

For case \emph{(i)}, we find the particle displacements throughout the lattice by substituting Eq.~(\ref{eq:eps2}) into Eqs.~(\ref{eq:av}) and solving the resulting partial differential equation: $\partial_y U_x + \partial_x U_y = - B x/2 $, with the additional constraint $\partial_x U_x = \partial_y U_y = 0$, which corresponds to nonvolumetric pure shear deformations. The resulting displacements satisfy
\begin{equation}
U_x = 0; \,\,\, U_y = - B x^2 /4.
\end{equation}
Note that for the honeycomb lattice, this condition can be realized using the boundary stresses illustrated in Fig.~\ref{Fig1}d.

For case  \emph{(ii)}, we substitute Eqs.~(\ref{eq:kappa}) into Eqs.~(\ref{eq:av}) to find the condition 
\begin{equation}
\label{eq:cond}
\sqrt{3} \partial_x  (\delta \kappa_2 - \delta \kappa_3)  - \partial_y (\delta \kappa_2 + \delta \kappa_3) = 3 \kappa B
\end{equation}
for the spatial dependence of the spring constants. We consider a material uniform along the $x$-direction. In this case, the condition in Eq.~(\ref{eq:cond}) is satisfied for spring constants given by
\begin{equation}
\label{eq:sel}
\alpha \equiv \frac{\delta \kappa_2}{\kappa} = \frac{\delta \kappa_3}{\kappa} = - \frac{\delta \kappa_1}{2\kappa} = \frac{B y}{3},
\end{equation}
which is visualized in Fig.~\ref{Fig1}e.
\begin{figure}
	\includegraphics[angle=0]{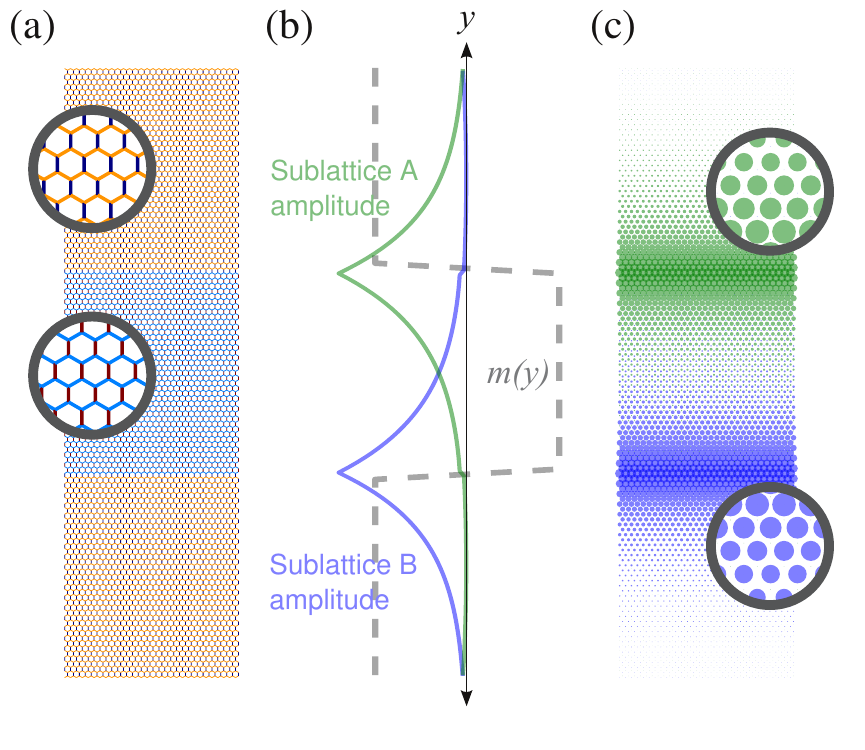}
	\caption{(a) Waveguide with two domain walls separating two regions with $\alpha = 0.08$ from a central region with $\alpha = -0.08$. The bonds are colored according to their spring constants as in Fig.~1(e). Periodic boundary conditions are applied along $x$. (b) Variation of the effective Dirac mass $m(y)$ (dashed line) and of the amplitude of the midgap mode at $q_x = 2 \pi / 3a$ on either sublattice (solid lines). (c) Visualization of the midgap mode with the sublattices distinguished. Each point is represented by a disc whose area is proportional to the amplitude of the midgap mode at that point. Points on sublattices A and B are drawn as green and blue discs respectively, showing the strong polarization of each domain wall mode onto a distinct sublattice.}
	\label{Fig3}
\end{figure}

\begin{figure}
	\includegraphics[angle=0]{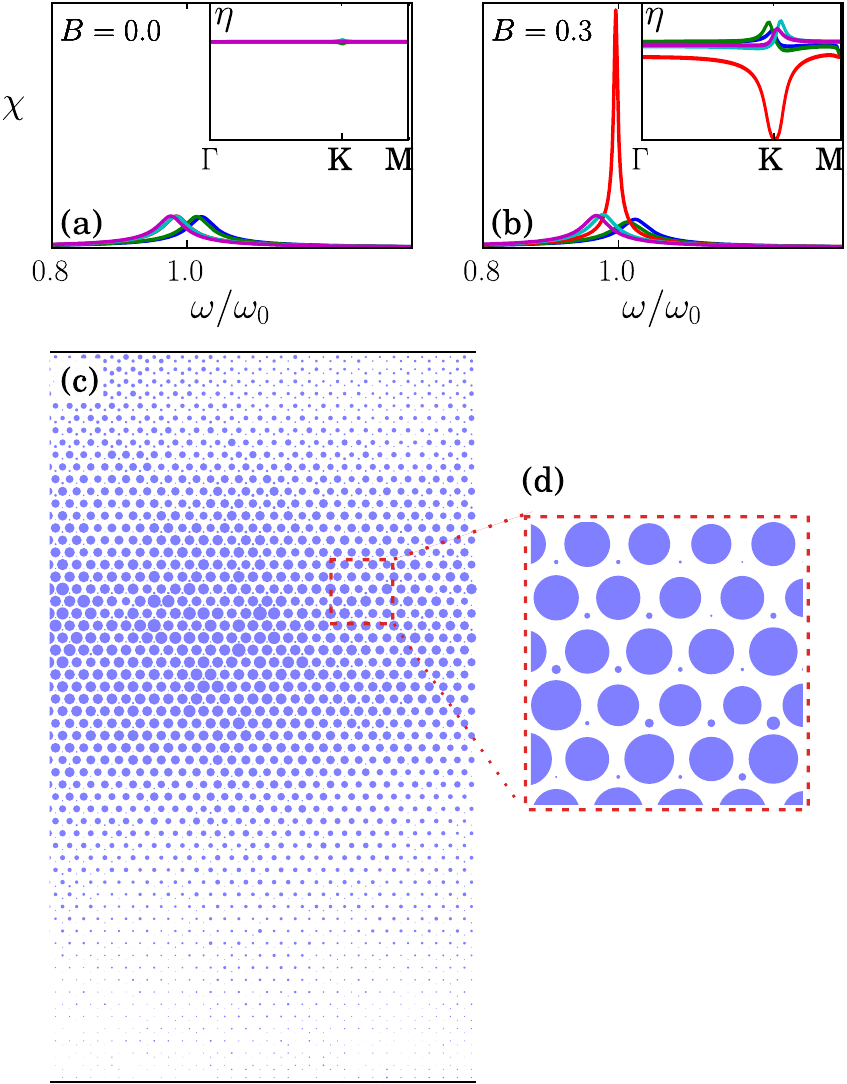}
	\caption{ Single-mode response $\chi$ of Landau-level states in mechanical graphene, including the effect of damping on one sublattice and for pseudo-magnetic field values (a) $ B = 0.0$ and (b) $B = 0.3$. Colors correspond to the different Landau-level bands identified in Fig.~2a. Insets: wavenumber-dependent attenuation rate $\eta$ of the corresponding bands. (c) The steady-state response (for $B = 0.3$) to  external periodic forcing with frequency close to the Dirac frequency and at an edge that is situated 50 unit cells to the left of the section shown.  Each point is represented by a disc whose area is proportional to the amplitude of the response. (d) Zoom-in of (c) shows that the Landau-level mode is selectively enhanced due to the presence of sublattice-biased damping.}
	\label{Fig4}
\end{figure}

{\it Mechanical Landau levels.}--- Now that we have proposed metamaterial architectures that realize the acoustic analog of a constant magnetic field, we go on to explore the physical consequences of this field for sound waves. To proceed, we focus on an architecture that is peculiar to the acoustic context, i.e., we select the realization of a patterned metamaterial waveguide described by Eqs.~(\ref{eq:sel}). Such a quasi-one-dimensional waveguide is uniform along the $x$-direction, graded along the $y$-direction, and is subject to no-stress boundary conditions on its top and bottom (see Fig.~\ref{Fig1}e). The constant pseudo-magnetic field leads to a Landau-level spectrum for frequencies near $\omega_0$ (Fig.~\ref{Fig2}a).

Let us focus on the acoustic band corresponding to the most prominent Landau level: $n=0$. In Fig.~\ref{Fig2}b, this band is plotted for several values of the pseudo-magnetic field $B$; as the pseudo-magnetic field increases, the band flattens over a larger region in wavevector space, which leads to an increasing peak in the density of acoustic states (shown in Fig.~\ref{Fig2}c). The width of this flat region defines an inverse length scale $\ell^{-1}$, which scales as $\ell^{-1} \sim \sqrt{B/a}$ (Fig.~\ref{Fig2}e). This length scale is the acoustic analog of the magnetic length of a Landau-level state~\cite{SI}. Consequently, an acoustic mode in the Landau level has a Gaussian profile with a transverse confinement given by $\ell$ (Fig.~\ref{Fig2}d). The transverse location of this mode within the waveguide is controlled by the mode wavenumber $q_x$, in contrast to an index-graded waveguide in which the location is determined by the mode frequency. In particular, modes with wavenumbers near the edge of the flat band are located near the waveguide edge, whereas modes near the center of the flat band are located in the waveguide bulk (Fig.~\ref{Fig2}d). 

{\it Sublattice-polarized domain wall modes.}--- The $n=0$ Landau level at $q_x=q_{K,x}\equiv 2\pi/3a$ has frequency $\omega_K$, is located at the waveguide center, and involves displacements exclusively on one sublattice. Modes with these properties generically appear in regions across which $A_x$ changes sign, i.e., their local dispersions have Dirac cones on opposite sides of point $K$. As an example, we consider a waveguide with two domain walls that separate a uniform central region with spring constants set by $\alpha = -0.08$ from two regions, one above and one below, that each have $\alpha = 0.08$ (Fig.~\ref{Fig3}a). At $q_x=q_{K,x}$, the spectrum as a function of $q_y$ near point $K$ is described by a gapped 1D Dirac Hamiltonian centered about $\omega_K$, with effective mass proportional to $A_x$~\cite{SI}. The ``spin'' degree of freedom corresponds to the two sublattices of the honeycomb lattice: eigenstates of $\sigma_z$ with eigenvalues $+1$ and $-1$ involve displacements solely on sublattice A and sublattice B, respectively. When the mass $m(y)$ varies spatially, domain walls at which $m(y)$ changes sign harbor exponentially localized midgap modes that are ``spin-polarized'', i.e., confined to a single sublattice~\cite{Jackiw1976,Su1979}. The sublattice on which the mode is localized is determined by the sign of the change in mass upon crossing the domain wall. Fig.~\ref{Fig3}b--c shows the numerically-obtained midgap mode for the domain wall geometry in Fig.~\ref{Fig3}a, whose components on sublattice A (sublattice B) fall off exponentially from the top (bottom) domain wall. 
 
{\it Towards mechanical lasers.}--- The sublattice polarization of the Landau-level states can also be used to selectively enhance these modes under external drive
by employing site-dependent damping. For example, for positive magnetic fields, the Landau-level states live only on the A-sublattice of the honeycomb unit cell~\cite{Jackiw1984, Guinea2010}.
If we introduce damping of the form $- \gamma \dot{\uv}_{\mathrm{B}}$ into the equation of motion, Eq.~(\ref{eq:motion}), such that only the displacements of the B-sublattice are damped, then the Landau-level acoustic waves would not be attenuated, whereas the rest of the sound waves, which generically are split between the A and B sublattices, would have a nonzero attenuation~\cite{SI}. 
To characterize this selective enhancement, we study the attenuation rate $\eta(\qv)$ as
a function of mode wavevector, as well as the self-response function $\chi(\omega)$ which measures the displacements in response to a drive at frequency $\omega$ (see~\cite{SI} for computation details). 
In Fig.~\ref{Fig4}a--b, we plot $\chi(\omega)$ and 
$\eta(\qv)$ for the Landau-level bands with $-2 \leq n \leq 2$, in response to an oscillatory drive that is proportional to the corresponding mode displacement vector.
In the absence of pseudo-magnetic field $B$, no mode stands out as having a strongest response peak in $\chi$~(Fig.~\ref{Fig4}a), whereas for nonzero $B$, 
$\chi$ exhibits a strong peak at a frequency $\omega_0$, corresponding to the zeroth Landau-level~(Fig.~\ref{Fig4}b).
Therefore, when an edge of the metamaterial is driven near $\omega_0$, the pseudo-magnetic field combined with selective damping leads to selective enhancement of acoustic Landau-level modes (Fig.~\ref{Fig4}c--d) relative to the rest of the attenuated acoustic spectrum.
This phenomenon is the acoustic analog of selective enhancement of microwave modes realized in Ref.~\cite{Poli2014}. 
Just as selective enhancement for light waves may lead to the design of novel parity-time-symmetric~\cite{Hodaei2014, Feng2014}
 and topological~\cite{Schomerus2013} lasers,
analogously, the selective enhancement of sound waves may be useful in the design of sound amplification by stimulated emission of radiation (SASERs), i.e., the acoustic analog of lasers, as well as acoustic couplers and rectifiers.

{\it Acknowledgments} ---  We gratefully acknowledge funding from FOM, NWO, and Delta Institute for theoretical physics (H.A., A.S., J.P., and V.V.)
and from EPSRC Programme Grant No. EP/N031776/1 (H.S.).

H.A. and A.S. contributed equally to this work.

\pagebreak

\renewcommand{\theequation}{S\arabic{equation}}

\section{Supplementary information for ``Synthetic gauge fields and Landau-level lasing in acoustic metamaterials''}
Here, we provide derivations for Eqs.~(4, 6) of the main text, the domain-wall-localized modes, and selective enhancement.

\subsection{Synthetic gauge fields for strain and patterning} 
In a periodic material, the equations of motion (1) admit plane-wave solutions $\uv = \uv_\qv e^{i (\omega(\qv) t - \qv\cdot \xv)}$
associated with the eigenvalue problem $ D(\qv) \uv_\qv = m \omega_\qv^2 \uv_\qv$, where
the dynamical matrix of the two-dimensional honeycomb lattice is:
\begin{equation}
\label{eq:Dmatrix}
\tilde{D}(\qv)=\frac{1}{m}\sum_\alpha \kappa_\alpha \left(\begin{array}{lcl}
P_\alpha & - P_\alpha e^{i\qv\cdot \delv_\alpha}\\
-P_\alpha e^{-i\qv \cdot \delv_\alpha}\ & P_\alpha \\
\end{array} \right),
\end{equation}
where $P_\alpha=\ev_\alpha \ev_\alpha^{\scriptscriptstyle T}$. Of the four bands of this dynamical matrix, two of them are degenerate at the Dirac point. Using first-order perturbation theory around the Dirac point, we find the following form for the dynamical matrix projected onto the two Dirac bands
\begin{equation}
D_0 = -\frac{3\kappa}{4m} a\delta\qv\cdot {\bm{{\rm \sigma}}} + \frac{3\kappa}{2m} \mathbb{I}.
\end{equation}
Using this approach, we introduce various perturbations. The deformation of the lattice are given by $ \Uv_{1,2} = \Uv \pm \Wv/2 $, where the different signs are used for the different sublattices. This deformation changes the components of the dynamical matrix via $\delv_{\alpha} \rightarrow \tilde{\delv}_{\alpha}$ and $P_\alpha \rightarrow \tilde{P}_\alpha$, where
\begin{align}
\tilde{\delv}_{\alpha} &= (\textit{I}+\nabla \Uv) \delv_{\alpha}  \\ 
\tilde{P}_\alpha&= P_\alpha+ (\nabla \Uv)P_\alpha+P_\alpha(\nabla \Uv)^{\scriptscriptstyle T}\nonumber\\
&+\left( \ev^{\scriptscriptstyle T}_\alpha \, \epsilon \, \ev_\alpha +\ev_{\alpha}^{\scriptscriptstyle T} \, \Wv/a\right) \left( \textit{I}-3P_\alpha\right)+\ev_\alpha \Wv^{\scriptscriptstyle T}/a+\Wv \, \ev_\alpha^{\scriptscriptstyle T}/a,
\end{align}
and $ (\nabla \Uv)_{ij} = \partial_i \Uv_j $.
Substituting these parameters into Eq.~(\ref{eq:Dmatrix}) and using perturbation theory, we obtain Eq.~(4) of the main text. 

In the main text, we discussed those configurations in which the lattice strain results from stress applied only to the boundaries. For such configurations, we imposed the force balance condition within the bulk of the lattice: $ \delta U_{tot} / \delta \Wv = 0 $, where $ U_{tot} $ is the potential energy associated with each unit cell and is given by
\begin{align}
U_{tot}(\Rv)= &U_1(\Rv,\Rv)+\frac{1}{2} \left[ U_2 (\Rv, \Rv - \av_2) +U_2(\Rv + \av_2 , \Rv)\right]\nonumber\\
&+\frac{1}{2} \left[ U_3 (\Rv , \Rv - \av_1) +U_3(\Rv + \av_1 , \Rv)\right],
\end{align}
where $ U_\alpha(\Rv_1,\Rv_2)=\frac{\kappa}{2}\left( \ev_\alpha \cdot\left[ \Uv_2 (\Rv_2) -\Uv_1(\Rv_1) \right] \right)^2 $. Solving the force-balance equation $ \delta U_{tot} / \delta \Wv = 0 $ using this energy, we get the result $ W_x=\epsilon_{xy}a$ and $ W_y=\frac{1}{2}(\epsilon_{xx}-\epsilon_{yy})a$ presented in the main text.

To obtain Eq.~(6) of the main text, we insert different spring constants for each of the three springs inside of each unit cell into Eq.~(\ref{eq:Dmatrix}) and keep the leading terms in the expansion.

\subsection{The sublattice-polarized modes} Some of our results for the sublattice-polarized modes can be understood using a connection between mechanical graphene and the Jackiw-Rebbi model~\cite{Jackiw1976}.
Note that, for the waveguide described in the main text, $\kappa_2=\kappa_3$ and $V =0$, which is equivalent to $\delta \kappa_2 = \delta \kappa_3 =-\delta\kappa_1/2 \equiv \kappa\alpha$. This case with staggered spring constants along the $y$-direction is reminiscent of the Su-Schrieffer-Heeger model~\cite{Su1979}. Using the formula for the synthetic gauge field in Eq.~(4) of the main text, we find the following form for the dynamical matrix:
\begin{equation}
D = D_0 + \frac{1}{3} \omega^2_0 \alpha(y) \sigma_x.
\end{equation}
Thus, we see that the dimensionless parameter $\alpha$ plays the role of the effective mass in the Jackiw-Rebbi model. For modes of the honeycomb-lattice waveguide near the Dirac frequency, we can obtain the form of the eigenmodes using the zero mode solution of the Jackiw-Rebbi model: $\uv(y) \propto \exp\left[ -\int_{0}^{y}\alpha(y)\,dy\right] $. Thus, for the sharp domain wall of Fig.~3, for which the effective mass is a step-function, we find solutions at the domain wall and which decay exponentially away from the domain wall. 
On the other hand, for a mechanical Landau-level mode obtained using material patterning, the mass varies linearly with $y$, i.e.,  $\alpha \propto y$, and the mode indeed
has a Gaussian profile. Furthermore, the solutions to the Jackiw-Rebbi model exhibit the parity anomaly, which can be used to ascertain that
 the domain-wall-bound modes as well as the Landau-level modes are both sublattice-polarized. 

\subsection{Selective enhancement} We use the drag matrix $\Gamma =\left(\begin{array}{cc}
\gamma \mathbb{I} &0 \\
0&\gamma' \mathbb{I} \end{array}\right)$ with $\gamma'=0$ to model sublattice-biased dissipation. With the presence of these drag forces, the equation of motion becomes $m\ddot{\uv} + \Gamma \dot{\uv} + D \uv = 0$. Now consider an external driving force $\Fv(\Rv, t) = \Fv(\Rv)e^{i \omega t}$ which oscillates in time. With this force, the inhomogenous equation of motion is $m\ddot{\uv}+\Gamma \dot{\uv} +D\uv = \Fv$. To find the solutions, we use Bloch functions, i.e., the normal modes of the periodic structure, to expand the drive as  $\Fv(\Rv, t) = e^{i \omega t} \sum_{n\kv} \Fv_{n\kv} \uv_{n\kv}e^{i\kv \cdot \Rv}$. A steady-state solution, if it exists, oscillates with the same frequency $\omega$ as the drive. The steady-state solution can be expanded as $\uv(\Rv, t) = e^{i\omega t} \sum_{\kv} c_{n\kv} u_{n\kv} e^{i\kv \cdot \Rv}$. From the equation of motion, we find that the coefficients $c_{n\kv}$ obey
\begin{equation}
c_{n\kv} = \frac{\Fv_{n\kv}}{-m \omega^2 + i \Gamma_{n\kv} \omega + \lambda_{n\kv}},
\end{equation}
where $\Gamma_{n\kv} = \sum_{n^\prime}\uv ^\dagger_{n^\prime\kv} \Gamma  \uv_{n\kv}$ and $\lambda_{n\kv}$ are the eigenvalues of the dynamical matrix, Eq.~(1) in the main text.
We expect the response to depend strongly on the damping. To see this, consider $\Gamma_{n\kv}$ to be a real number. If $\Gamma_{n\kv}^2 > 2\lambda_{n\kv}m$ (corresponding to the overdamped limit), then the amplitude of the response never rises above $\Fv_{n\kv}/\lambda_{n\kv}$ -- it attains this limiting value at low frequencies and falls off as $\Fv/m\omega^2$ at higher frequencies. If on the other hand, $\Gamma_{n\kv}^2 < 2\lambda_{n\kv}$ (corresponding to the underdamped limit), the response develops a peak at $\omega^2_p = \lambda_{n\kv}/m - \Gamma_{n\kv}^2/2m^2$, whose height diverges as $1/\Gamma_{n\kv}$. Therefore, at low damping, the response will be dominated by modes whose natural frequency is close to the driving frequency. If for example, the lattice is driven by forcing atoms along one edge in an oscillatory manner, then $\Fv_{n\kv}$ will be appreciable for several modes, but the only modes to have a strong response will be those whose natural frequency is close to the driving frequency. 

This observation can be used to selectively enhance the zeroth Landau level mode, as seen in Fig.~4 of the main text.
In Fig.~4a-b, we plot $\chi(\omega) \equiv c_{n\kv}/F_{n\kv}$
for $n$ corresponding to Landau levels $-2,-1,0,1,$ and $2$,
for $\kv = \qv_K$.
In the insets, we plot the attenuation rates $\eta$, corresponding
to the imaginary parts of the frequency spectrum, for these modes
as a function of $\kv$ along the $\Gamma K M$ cut of the Brillouin zone. 
For both quantities, the zeroth Landau level mode is selected for nonzero $B$:
it has a stronger response and 
smaller attenuation than the other modes.
We then drive the lattice with force $\Fv(t) = e^{i \omega t } \hat{x}$ ($\omega$ near $\omega_0$),
on two of the lattice points (near but slightly above the waveguide center), 
and observe the amplitude of the steady-state response sufficiently far
away from this drive. We note that the Landau-level mode is selectively enhanced.

\end{document}